\documentclass[journal=nalefd,manuscript=article]{achemso}
\usepackage{chemformula} 
\usepackage[T1]{fontenc} 
\usepackage[version=3]{mhchem}

\usepackage{graphicx}
\usepackage{dcolumn}
\usepackage{bm}

\usepackage[utf8]{inputenc}
\usepackage[T1]{fontenc}
\usepackage{etoolbox}

\usepackage{color}

\usepackage{amssymb}
\usepackage{amsmath}

\usepackage{subeqnarray}

\usepackage{color}

\title[]{Second-order topology in two-dimensional azulenoid kekulene carbon lattices}

\author{Xiaorong Zou}
\affiliation{Center for 2D Quantum Heterostructures (2DQH), Institute for Basic Science (IBS), Suwon 16419, Republic of Korea}

\author{Hyeon Suk Shin}
\affiliation{Center for 2D Quantum Heterostructures (2DQH), Institute for Basic Science (IBS), Suwon 16419, Republic of Korea}
\alsoaffiliation{Department of Chemistry, Sungkyunkwan University, Seobu-ro 2066, Suwon 16419, Republic of Korea}
\alsoaffiliation{Department of Energy, Sungkyunkwan University, Seobu-ro 2066, Suwon 16419, Republic of Korea}

\author{Chang-Jong Kang}
\affiliation{Department of Physics, Chungnam National University, Daejeon 34134, Republic of Korea}
\alsoaffiliation{Institute for Sciences of the Universe, Chungnam National University, Daejeon 34134, Republic of Korea}

\author{Baibiao Huang}
\affiliation {School of Physics, State Key Laboratory of Crystal Materials, Shandong University, Jinan 250100, China}

\author{Yanmei Zang}
\affiliation{Department of Energy Science, Sungkyunkwan University, Seobu-ro 2066, Suwon 16419, Republic of Korea}
\alsoaffiliation{Department of Energy, Sungkyunkwan University, Seobu-ro 2066, Suwon 16419, Republic of Korea}

\author{Ying Dai}
\affiliation {School of Physics, State Key Laboratory of Crystal Materials, Shandong University, Jinan 250100, China}
\email{daiy60@sdu.edu.cn}
\author{Chengwang Niu}
\affiliation {School of Physics, State Key Laboratory of Crystal Materials, Shandong University, Jinan 250100, China}
\email{c.niu@sdu.edu.cn}
\affiliation {School of Physics, State Key Laboratory of Crystal Materials, Shandong University, Jinan 250100, China}

\author{Chang Woo Myung}
\email{cwmyung@skku.edu}
\affiliation{Center for 2D Quantum Heterostructures (2DQH), Institute for Basic Science (IBS), Suwon 16419, Republic of Korea}
\alsoaffiliation{Department of Energy Science, Sungkyunkwan University, Seobu-ro 2066, Suwon 16419, Republic of Korea}
\alsoaffiliation{Department of Energy, Sungkyunkwan University, Seobu-ro 2066, Suwon 16419, Republic of Korea}
\alsoaffiliation{Department of Quantum Information Engineering, Sungkyunkwan University, Seobu-ro 2066, Suwon, 16419, Korea}

\date{\today}

\title[An \textsf{achemso} demo]
{Second-order topology in two-dimensional azulenoid kekulene carbon lattices}


\begin{document}

	\begin{abstract}
		The discovery of higher-order topological insulator (HOTI) has established a new paradigm for understanding symmetry-constrained boundary electronic states. Here, based on first-principles calculations, we demonstrate the emergence of HOTI phase in organic lattices of two-dimensional azulenoid–kekulene-type carbon allotropes, namely AKC-[3,3] and AKC-[6,0]. Enabled by the $\mathcal{C}_6$ rotational symmetry, the nontrivial bulk topology is confirmed through the topological invariant and fractionally quantized corner charge, giving $\{[M^{(I)}_{2}],[K^{(3)}_{2}]\}$ = $\{0,2\}$ and $Q_{\mathrm{corner}} = e/3$, respectively, accompanied by the emergence of exotic corner states in nanoflakes. Notably, the structural modifications are explored, revealing that in the derived structure PAK-[6,0], whose corner-localized states are preserved, highlighting the robustness of the higher-order topological phase. These findings highlight azulenoid–kekulene-based carbon allotropes as a promising platform to explore the interplay between structural design, crystalline symmetry, and higher-order topological boundary responses in two-dimensional carbon systems.
	\end{abstract}

	\maketitle
	
	Two-dimensional (2D) carbon allotropes are a key class of low-dimensional materials, owing to their structural versatility and chemical tunability, which make them attractive for applications in nanoelectronics and energy storage, as well as for exploring novel electronic properties~\cite{2008sciencecarbon,2012CSRCARBON,2013sciencecarbon,2019CSRcarbon,2021sciencecarbon,2021RSERCARBON}. Many studies mainly focused on graphene-based lattices composed of hexagonal rings, where lattice symmetry gives rise to well-understood electronic properties~\cite{Kane,2009RMPCARBON,2009sciencecarbon,2014PRLcarbon,2019AMcarbon,2019NRPcarbon,2022NCcarbon}. However, such systems are largely built from a single type of structural unit, which limits the diversity of achievable lattice configurations and electronic behaviors. Consequently, attention has extended to carbon allotropes that incorporate non-hexagonal rings, such as pentagons and heptagons~\cite{2015NLcarbon,2020jacsCARBON,2021ACScarbon,2022AScarbon}. The inclusion of these non-hexagonal units significantly enriches structural complexity at the unit-cell scale,
	modifying atomic connectivity and inducing multi-orbital hybridization of carbon $p$-orbitals,
	which offers tremendous potential for creating rich band topologies~\cite{2019carbon,2020carbon,2021PRBcarbon,2021JPCLcarbon,2025PRBcarbon}. Recently, a new class of two-dimensional carbon allotropes based on azulenoid–kekulene building blocks has been proposed~\cite{2024NCcarbon}. These azulenoid–kekulene carbon (AKC) lattices, constructed from such non-hexagonal units, have been demonstrated to be energetically favorable and thermally stable among a wide range of proposed 2D carbon allotropes, while preserving long-range crystalline order. However, the electronic properties of such systems remain to be systematically investigated.
	
	Higher-order topological insulators (HOTIs) extend the conventional bulk–boundary correspondence~\cite{highorderfirst,2021NRPn12,2021NRPn1} by supporting lower-dimensional boundary states, including corner states in two dimensions and hinge or corner states in three dimensions~\cite{multipolescience,highordersonginvariants,highorderezawa1,highorderrenyafei,Peterson1114}. These phases, typically protected by crystalline symmetries~\cite{2021PRBzengjiang,2021prbyaoyugui}, have been extensively studied, with a wide range of fundamental phenomena continuously proposed and systematically explored~\cite{Hsu13255,ZhangDPRL2019,2019PRLhotisc,2022NCHOTI,2022npjfese,2024LSAhoti,2025NChoti,Bai2026PRL}. The HOTIs can emerge in purely carbon-based two-dimensional allotropes, such as graphdiyne, and, notably, can be further extended to organic systems~\cite{highordersplit,highordergranpj,highersw,2022jacsOhoti,2022JPCLhoti,2023MTNhoti,hoti2d}.
	To date, most studies have concentrated on systems with simple lattice geometries and small primitive cells~\cite{higherordereuln2as2,2020NMWTe2,highorderbieuo}. By contrast, two-dimensional lattices incorporating non-hexagonal rings and complex real-space connectivity, as frequently encountered in realistic materials, have received little attention in the context of higher-order topology. This calls for a systematic investigation of the emergence and characteristics of HOTIs in these systems. 
	
	In this work, we employ first-principles calculations to systematically investigate the topological properties of two representative azulenoid–kekulene carbon allotropes, AKC-[3,3] and AKC-[6,0]. Both systems are found to be bulk insulators with finite energy gaps,  Remarkably, symmetry analysis reveals a nonzero topological invariant $\{[M^{(I)}_{2}],[K^{(3)}_{2}]\}$ = $\{0,2\}$, along with fractionally quantized corner charges of $e/3$, confirming the HOTI phase in AKC-[3,3] and AKC-[6,0]. Moreover, finite hexagonal nanoflakes that preserve the intrinsic $\mathcal{C}_6$ rotational symmetry are constructed, in which well-defined in-gap states appear within the bulk gap, with charge density strongly localized at the six corners, further explicitly demonstrating the nontrivial higher-order topology. In addition, the higher-order topological characteristics are preserved in PAK-[6,0], demonstrating the robustness against structural modification. Our results suggest that higher-order topology can persist in two-dimensional carbon lattices with non-hexagonal rings and enlarged unit cells, extending its realization to a wider range of lattice geometries.
	
	First-principles calculations are carried out within the framework of density functional theory as implemented in the Vienna Ab initio Simulation Package (VASP)~\cite{Kresse,Kresse1}. The exchange–correlation interaction is described using the generalized gradient approximation in the Perdew–Burke–Ernzerhof (PBE) form~\cite{pbevasp}.
	To eliminate spurious interactions between periodic images, a vacuum spacing of 20 \AA{} is introduced along the out-of-plane direction. All atomic positions are fully optimized until the residual forces on each atom below 0.01 eV/\AA ~, with a plane-wave kinetic-energy cutoff of 500 eV. The electronic self-consistency criterion is set to 10$^{-6}$ eV. Maximally localized Wannier functions (MLWFs) are subsequently constructed using the Wannier90 package~\cite{wannier90}, and the edge states are further calculated based on the WannierTools~\cite{WU2017}.
	
	\begin{figure} 
		\centering
		\includegraphics[width=1\linewidth]{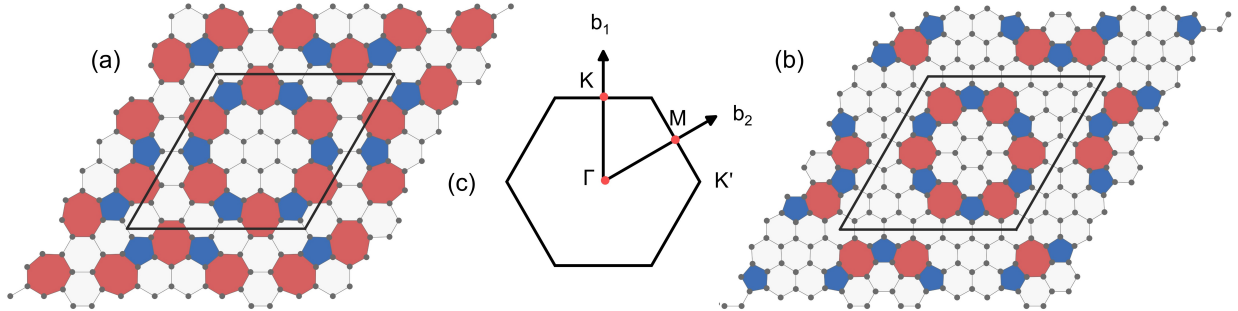}
		\caption{Top views of (a) AKC-[3,3] and (b) AKC-[6,0] lattices. The unit cells are outlined by black lines. (c) Schematic illustration of the hexagonal Brillouin zone with high-symmetry points. The reciprocal lattice vectors are denoted by $b_1$ and $b_2$.}
		\label{structure}
	\end{figure}
	
	AKC materials are constructed by periodically incorporating azulenoid–kekulene(AK) units into a graphene-derived hexagonal framework, forming two-dimensional crystalline structures with a hexagonal Bravais lattice~\cite{2024NCcarbon}. By introducing fused pentagon–heptagon motifs into the hexagonal network, the AKC structures exhibit significantly enlarged primitive cells while preserving the overall lattice symmetry. The differences between various AKC allotropes primarily arise from the periodic arrangement of AK units within the parent lattice, leading to distinct spatial periodicities and local geometric environments under the same symmetry constraints.
	
	The two allotropes considered in this work, AKC-[3,3] and AKC-[6,0], correspond to different superlattice parameters determined by the relative arrangement of AK units. In AKC-[3,3] shown in Fig.~\ref{structure}(a), the AK units are more densely packed, with neighboring units interconnected through extended $\pi $-conjugated networks. In contrast, AKC-[6,0], shown in Fig.~\ref{structure}(b), features more widely separated AK units within an extended hexagonal carbon framework incorporating pentagon–heptagon motifs. The structural differences are also manifested in the primitive cell size, with 54 and 72 carbon atoms per unit cell for AKC-[3,3] and AKC-[6,0], respectively. Despite their distinct bonding configurations and geometric scales, both structures belong to space group No. 191 (P6/mmm) and preserve inversion symmetry $\mathcal P$ as well as sixfold rotational symmetry $\mathcal C_6$. 
	
	\begin{figure}
		\centering
		\includegraphics[width=0.6\linewidth]{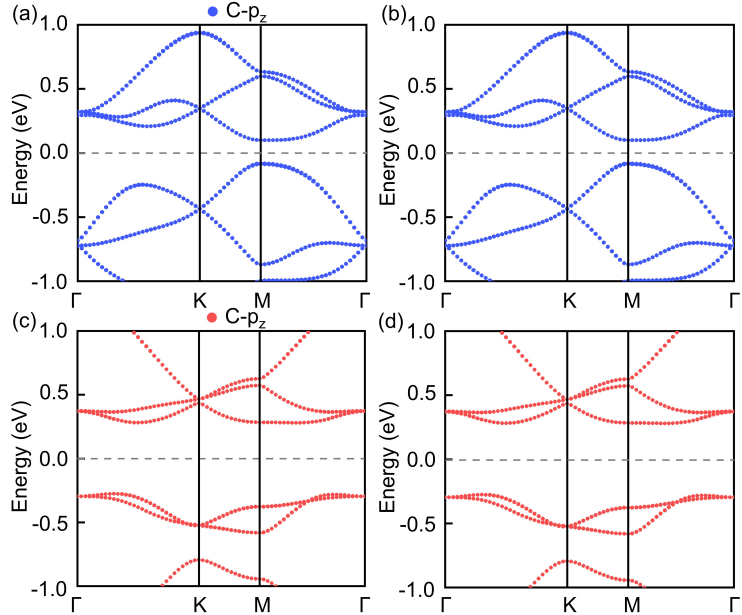}
		\caption{Electronic band structures of AKC-[3,3] lattice (a) without and (b) with spin–orbit coupling (SOC). Band structures of AKC-[6,0] lattice (c) without and (d) with SOC.} 
		\label{band}
	\end{figure}
	
\begin{figure}
		\centering
		\includegraphics[width=0.6\linewidth]{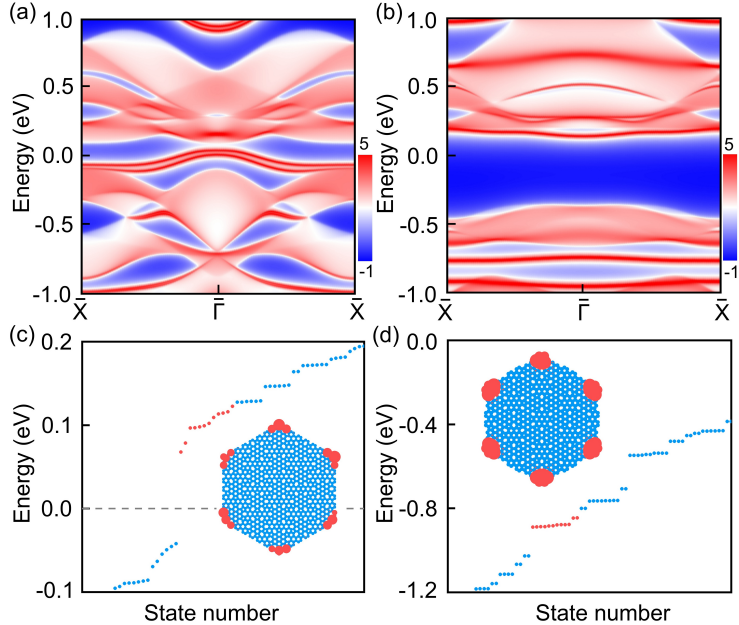}
		\caption{Edge-state spectra of (a) AKC-[3,3] and (b) AKC-[6,0] lattices. Discrete energy spectra for finite (c) AKC-[3,3] and (d) AKC-[6,0] nanoflakes. The orange dots indicate the in-gap corner states. Insets show the corresponding real-space charge distributions of the corner states, as marked by orange dots.} 
		\label{hoti}
	\end{figure}

	Having established the structural features and symmetry constraints, we next examine the electronic structure of AKC-[3,3]. First-principles calculations indicate that the band dispersions with and without SOC are nearly unchanged around the Fermi level, as illustrated in Figs.~\ref{band}(a) and~\ref{band}(b). The system is thus identified as a 2D insulator with a bulk gap of approximately 0.18 eV. To characterize its topological nature, we evaluate the edge states using the Green’s function formalism. As shown in Fig.~\ref{hoti}(a), although the edge-related states emerge within the bulk gap, the edge spectrum remains fully gapped, with no modes connecting the conduction and valence bands. The presence of a finite bulk gap together with the absence of gap-crossing edge modes rules out conventional first-order topology in AKC-[3,3], but instead signifies a higher-order topological phase. To confirm the higher-order topological character, we construct a finite hexagonal nanoflake that preserves the intrinsic $\mathcal C_6$ rotational symmetry. The calculated energy spectrum displays twelve in-gap states, as illustrated in Fig.~\ref{hoti}(c). Analysis of the corresponding charge density distributions reveals that these states are predominantly localized at the six corners of the nanoflake, with negligible weight along the edges and in the bulk region. The emergence of symmetry-related corner states therefore provides direct real-space evidence that AKC-[3,3] realizes a second-order topological insulating phase. To further verify the bulk origin of the corner states, we construct an alternative nanoflake with a different edge termination. The corner-localized in-gap states remain robust in Fig. S1, indicating that they are insensitive to boundary modification and originate from the underlying bulk topology.
	
	In addition, we compute symmetry-based topological indicators defined by crystalline symmetry constraints. These invariants, obtained from the symmetry representations of the occupied electronic bands at high-symmetry points in the Brillouin zone, provide a bulk characterization of higher-order topology~\cite{c3invariant,highorderinvariants,higher2020invariant}. For a crystal possessing $\mathcal C_n$ rotational symmetry, the eigenvalues of the rotation operator at a high-symmetry momentum point $\Pi$ take the form
	\begin{align}
		\lambda_p^{(n)} = e^{i \frac{2\pi (p-1/2)}{n}}, \quad p=1,\dots,n .
	\end{align}
	Based on these symmetry eigenvalues, symmetry indicators can be constructed by comparing the number of occupied bands associated with each rotation eigenvalue at different high-symmetry points relative to the reference point $\Gamma$. The corresponding quantity is defined as
	\begin{align}
		[\Pi_p^{(n)}] = N_{\Pi,p} - N_{\Gamma,p}
	\end{align}
	where $N_{\Pi,p}$ denotes the number of occupied bands at momentum $\Pi$ carrying the rotation eigenvalue $\lambda_p^{(n)}$. These indicators encode the symmetry information of the bulk electronic structure and enable the diagnosis of nontrivial topological phases protected by crystalline rotational symmetry. For two-dimensional insulators with $\mathcal C_6$  symmetry, the fractional corner charge can be determined from the symmetry indicators associated with the $M$ and $K$ points according to
	\begin{align}
		Q_{\mathrm{corner}} = \frac{e}{4}[M_2^{(2)}] + \frac{e}{6}[K_2^{(3)}].
	\end{align}
	
	If inversion symmetry $P$ is additionally preserved, the indicator $[M_2^{(2)}]$ can be equivalently expressed in terms of the inversion eigenvalue, and the above relation can be rewritten as
	\begin{align}
		Q_{\mathrm{corner}} = \frac{e}{4}[M_2^{(I)}] + \frac{e}{6}[K_2^{(3)}].
	\end{align}
	Here, $M_2^{(I)}$ represents the difference in the number of occupied bands with inversion eigenvalue -1 between the $M$ and $\Gamma$ points, and
	$K_1^{(3)}$ denotes the difference in the number of occupied bands carrying the $\mathcal C_3$ rotational eigenvalue -1 between the $K$ and $\Gamma$ points. For the AKC-[3,3] lattice, the calculated indicators are $\{[M^{(I)}_{2}],[K^{(3)}_{2}]\}$ = $\{0,2\}$.
	Substituting these values into the above relation yields $Q_{\mathrm{corner}} = e/3$, indicating the accumulation of a fractionally quantized charge of $e/3$ at each corner of a hexagonal nanoflake. This quantized corner charge is consistent with the presence of corner-localized in-gap states and therefore provides an independent topological characterization of the second-order insulating phase realized in AKC-[3,3].
	
	Having established the second-order topological character of AKC-[3,3], we next examine AKC-[6,0], which represents a distinct structural realization under the same crystallographic symmetry constraints. Band structure calculations show that AKC-[6,0] is a two-dimensional insulator with an enlarged bulk gap of approximately 0.55 eV, as shown in Fig.~\ref{band}(d). The edge-states spectrum remains fully gapped within the bulk energy window shown in Fig.~\ref{hoti}(b), implying that AKC-[6,0] belongs to the same higher-order topological class as AKC-[3,3]. And the calculations on finite-size hexagonal nanoflakes further reveal in-gap states inside the bulk gap. The corresponding charge density distributions exhibit strong localization at the six corners of the nanoflake, as illustrated in Fig.~\ref{hoti}(d). Symmetry-indicator analysis under $\mathcal C_6$ rotational symmetry yields a fractionally quantized corner charge of e/3, placing AKC-[6,0] in the same topological category as AKC-[3,3]. The consistency of the symmetry indicators and corner charge demonstrates that the higher-order topological phase in these AKC lattices is constrained by crystalline symmetry rather than by specific geometric scale. In this sense, the topological classification is determined  by symmetry eigenvalues at high-symmetry momenta. A similar robustness against edge modification is observed for AKC-[6,0] shown in Fig. S2, where the corner-localized states persist under different edge termination.
	
	\begin{figure}
		\centering
		\includegraphics[width=0.6\linewidth]{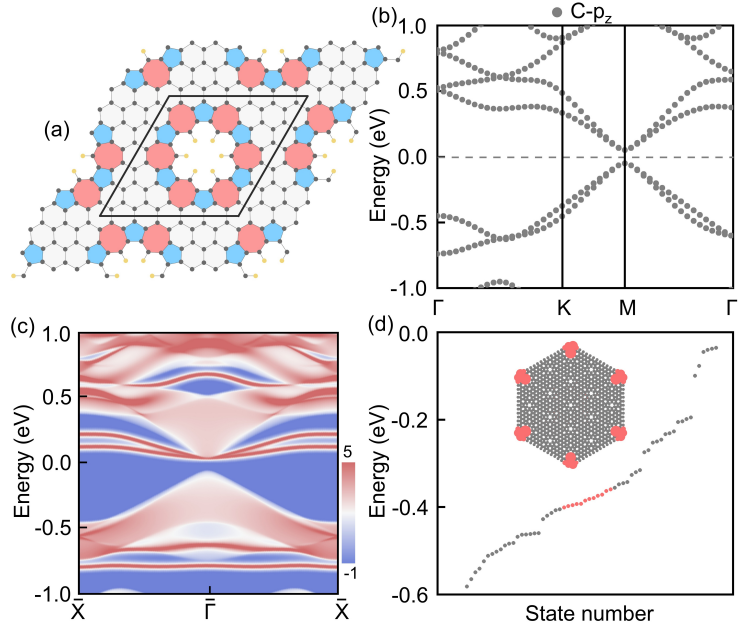}
		\caption{(a) Top view of the crystal structure for PAK-[6,0]. Gray and yellow spheres denote C and H atoms, respectively. (b) Electronic band structure and (c) edge states of PAK-[6,0]. (d) Discrete energy spectra for a finite PAK-[6,0] nanoflake. The pale orange dots indicate the in-gap corner states. Inset shows the corresponding real-space charge distributions of the corner states, as marked by pale orange dots.} 
		\label{pak}
	\end{figure}
	
	In realistic materials, structural modulations and variations in the chemical environment are often unavoidable, such as atomic vacancies and hydrogen passivation. It is therefore of particular interest to examine whether higher-order topological phases remain robust under such perturbations. To this end, we consider the porous azulenoid–kekulene structure PAK-[6,0], which can be regarded as a structurally perturbed counterpart of AKC-[6,0]. Compared with AKC-[6,0], the PAK-[6,0] structure exhibits a porous geometry with hydrogen-passivated boundaries, as shown in Fig.~\ref{pak}(a), leading to noticeable changes in the local bonding environment and atomic coordination. Despite these structural modifications, the overall lattice symmetry is still preserved. The results show that the edge states of PAK-[6,0] remain gapped within the bulk energy gap,, as illustrated in Fig.~\ref{pak}(c). Meanwhile, in the corresponding hexagonal nanoflake, a set of corner-localized states can be identified inside the bulk gap in Fig.~\ref{pak}(d), and the associated corner charge remains fractionally quantized, consistent with that of the AKC systems. These results indicate that, despite the introduction of local structural reconstruction, the key signatures of the higher-order topological phase are still preserved, demonstrating the robustness of the higher-order topology. 
	
	In summary, we have established that both AKC-[3,3] and AKC-[6,0] realize the HOTI phase under the same crystalline symmetry constraints. This classification is consistently manifested in fully gapped edge dispersions, symmetry-related corner-localized states, and fractionally quantized corner charges. By comparing structurally distinct realizations, we show that the higher-order topological phase in these AKC lattices are dictated mainly by crystalline symmetry, rather than primitive cell size, and is robust against structural variations. Our results provide a concrete realization of higher-order topology in two-dimensional carbon lattices with non-hexagonal rings and complex connectivity, and offer a useful platform for exploring topological phases in carbon-based systems. The geometric isolation and topological protection of the resulting zero-dimensional states indicate their robustness against perturbations, suggesting potential relevance for nanoscale electronic and quantum applications.
	
	\medskip
	\textbf{Acknowledgements} \par 
	This work was supported by Institute for Basic Science (IBS-R036-D1) and the Taishan Scholar Program of Shandong Province. We are grateful for the computational support from the Korea Institute of Science and Technology Information (KISTI) (KSC-2023-CRE-0355, KSC-2023-CRE-0261, KSC-2023-CRE-0502, KSC-2024-CRE-0144, KSC-2024-CRE-0088, KSC-2024-CRE-0117). Computational work for this research was partially performed on the Olaf supercomputer supported by IBS Research Solution Center and GPU cluster supported by Ministry of Science and ICT (MSIT) and the National IT Industry Promotion Agency (NIPA).
	
	
	\providecommand{\latin}[1]{#1}
	\makeatletter
	\providecommand{\doi}
	{\begingroup\let\do\@makeother\dospecials
		\catcode`\{=1 \catcode`\}=2 \doi@aux}
	\providecommand{\doi@aux}[1]{\endgroup\texttt{#1}}
	\makeatother
	\providecommand*\mcitethebibliography{\thebibliography}
	\csname @ifundefined\endcsname{endmcitethebibliography}
	{\let\endmcitethebibliography\endthebibliography}{}

\end{document}